\documentclass[11pt, epsf]{article}

\usepackage{epsfig}
\usepackage{amssymb}
\usepackage{graphicx}
\usepackage{color}
\usepackage{subfigure}
\usepackage{dsfont}

\begin{document}

\baselineskip 6mm
\renewcommand{\thefootnote}{\fnsymbol{footnote}}


\newcommand{\nc}{\newcommand}
\newcommand{\rnc}{\renewcommand}



\newcommand{\tcb}{\textcolor{blue}}
\newcommand{\tcr}{\textcolor{red}}
\newcommand{\tcg}{\textcolor{green}}


\def\be{\begin{equation}}
\def\ee{\end{equation}}
\def\ba{\begin{array}}
\def\ea{\end{array}}
\def\bea{\begin{eqnarray}}
\def\eea{\end{eqnarray}}
\def\nn{\nonumber\\}


\def\ct{\cite}
\def\la{\label}
\def\eq#1{(\ref{#1})}


\def\a{\alpha}
\def\b{\beta}
\def\g{\gamma}
\def\G{\Gamma}
\def\d{\delta}
\def\D{\Delta}
\def\e{\epsilon}
\def\et{\eta}
\def\ph{\phi}
\def\Ph{\Phi}
\def\ps{\psi}
\def\Ps{\Psi}
\def\k{\kappa}
\def\l{\lambda}
\def\L{\Lambda}
\def\m{\mu}
\def\n{\nu}
\def\th{\theta}
\def\Th{\Theta}
\def\r{\rho}
\def\s{\sigma}
\def\S{\Sigma}
\def\ta{\tau}
\def\o{\omega}
\def\O{\Omega}
\def\pr{\prime}


\def\half{\frac{1}{2}}

\def\goto{\rightarrow}

\def\na{\nabla}
\def\grad{\nabla}
\def\curl{\nabla\times}
\def\div{\nabla\cdot}
\def\pa{\partial}
\def\fr{\frac}

\def\bra{\left\langle}
\def\ket{\right\rangle}
\def\lb{\left[}
\def\lc{\left\{}
\def\ls{\left(}
\def\lp{\left.}
\def\rp{\right.}
\def\rb{\right]}
\def\rc{\right\}}
\def\rs{\right)}

\def\vac#1{\mid #1 \rangle}


\def\td#1{\tilde{#1}}
\def\check{ \maltese {\bf Check!}}


\def\Tr{{\rm Tr}\,}
\def\det{{\rm det}}


\def\bc#1{\nnindent {\bf $\bullet$ #1} \\ }
\def\ch {$<Check!>$ }
\def\ss {\vspace{1.5cm}}
\def\text#1{{\rm #1}}
\def\Id{\mathds{1}}

\begin{titlepage}

\hfill\parbox{5cm} {APCTP Pre2016 - 001 }

\vspace{25mm}

\begin{center}
{\Large \bf Logarithmic Corrections to the Entanglement Entropy }

\vskip 1. cm
  {Chanyong Park\footnote{e-mail : chanyong.park@apctp.org}}

\vskip 0.5cm

{\it  Center for Quantum Spacetime, Sogang University, Seoul 121-742, Korea}\\
{\it  Asia Pacific Center for Theoretical Physics, Pohang, 790-784, Korea } \\
{\it  Department of Physics, Postech, Pohang, 790-784, Korea }\\

\end{center}

\thispagestyle{empty}

\vskip2cm


\centerline{\bf ABSTRACT} \vskip 4mm

In a $d$-dimensional conformal field theory, it has been known that a relevant deformation operator with the conformal dimension, $\Delta=\frac{d+2}{2}$, generates a logarithmic correction to the entanglement entropy. In the large 't Hooft coupling limit, we can investigate such a logarithmic correction holographically by deforming an AdS space with a massive scalar field dual to the operator with $\Delta=\frac{d+2}{2}$.  There are two sources generating the logarithmic correction. One is the metric deformation and the other is the minimal surface deformation. In this work, we investigate the change of the entanglement entropy caused by the minimal surface deformation and find that the second order minimal surface deformation leads to an additional logarithmic correction.

\vspace{1cm}

\vspace{2cm}


\end{titlepage}

\renewcommand{\thefootnote}{\arabic{footnote}}
\setcounter{footnote}{0}


\section{Introduction}

The entanglement entropy is an important quantity as an order parameter of the quantum phase transition and describes the entanglement between quantum states. Recently, much attention has been paid to account for various quantum aspects of the condensed matter system as well as of the dual field theory in the AdS/CFT context. Although the entanglement entropy is well defined in the quantum field theory (QFT), it is usually formidable to calculate the entanglement entropy of an interacting QFT. In this situation, the AdS/CFT correspondence \cite{Maldacena:1997re,Gubser:1998bc,Witten:1998qj,Witten:1998zw,Aharony:2008ug} can shed light on understanding the entanglement entropy of a strongly interacting system via a relatively simple dual gravity calculation. 

For a two-dimensional conformal field theory (CFT), the exact entanglement entropy has been known due to the conformal symmetry and modular invariance \cite{Calabrese:2004eu,Calabrese:2005zw,Calabrese:2009qy}. In \cite{Ryu:2006bv,Ryu:2006ef,Nishioka:2009un}, authors have proposed how to calculate the entanglement entropy holographically in the dual gravity, which is usually called the holographic entanglement entropy. Similar to the black hole entropy, the holographic entanglement entropy is geometrized as the area of the minimal surface ending on the entangling surface 
\be
S = \fr{2 \pi A}{\k^2} ,
\ee
where $A$ denotes the area of the minimal surface. It has been proven that the holographic entanglement entropy can exactly reproduce the results obtained in the two-dimensional CFT \cite{Ryu:2006bv,Ryu:2006ef}. In the AdS/CFT context, a strongly interacting CFT is dual to a weakly curved gravity so that one can easily apply the holographic method to higher dimensional interacting CFTs. As a result, the AdS/CFT correspondence can alleviate our labors to calculate the entanglement entropy \cite{Solodukhin:2008dh}-\cite{Park:2015afa}. The holographic analysis has shown that, when one considers a subsystem in a disk-shaped region, the $A$-type central charge appears as the coefficient of the logarithmic term. In this case, the logarithmic term is universal in that it is independent of the regularization scheme \cite{Ryu:2006bv,Ryu:2006ef}. However, its existence crucially relies on the dimension and shape of the entangling surface.

In \cite{Hertzberg:2010uv}, it has been argued that there can exist another logarithmic correction when one deforms a $d$-dimensional CFT by the relevant operator with a specific conformal dimension 
\be
\D = \fr{d+2}{2} .
\ee  
Unlike the aforementioned logarithmic term related to the central charge, the additional logarithmic correction appears regardless of the shape of the entangling surface and is proportional to the entangling surface area \cite{Huerta:2011qi,Lewkowycz:2012qr}. From the perturbation calculation \cite{Rosenhaus:2014woa,Rosenhaus:2014ula,Rosenhaus:2014zza,Faulkner:2014jva}, those facts have been checked in a four-dimensional free massless fermion theory deformed by a fermion mass term \cite{Rosenhaus:2014zza}. In this case, the additional logarithmic correction occurs at the second order perturbation. This is the story of a free CFT. Although the conformal perturbation theory is well defined even in an interacting CFT, in practice it is not easy to calculate the entanglement entropy because of nontrivial quantum corrections. So it is interesting to ask whether the similar additional logarithmic correction occurs in a strongly interacting CFT. The holographic method enables us to investigate such a logarithmic correction even in the strong coupling regime \cite{Hung:2011ta}.

In general, a relevant operator deforms IR physics of the CFT. On the dual gravity side, such a deformation is realized as the modification of an inner geometry which also alters the area of the minimal surface.
In \cite{Hung:2011ta}, authors have investigated the logarithmic correction caused by the metric deformation in detail. They have found the general form of the logarithmic term caused by the metric deformation. In addition, they also showed 
that a more general relevant operator with $\D=\fr{(2n-1)d+2}{2 n}$, where $n$ denotes an integer number, can lead
to a more general logarithmic term 
\be
\d S \sim \l^{2n} A_{\S} \log \ls \l^{2n/(d-2)} \e\rs .
\ee
Note that in a free theory composed of bosons and fermions there is no such operator with the above conformal dimension for $n>2$. Regarding an interacting CFT, however, the conformal dimension can be shifted due to quantum corrections. For magnon's and spike's solutions having large conformal dimensions, their anomalous dimensions have been studied in numerous background geometries \cite{Minahan:2002ve}-\cite{Lee:2011fe}. Therefore, it seems to be natural in an interacting QFT to take into account a quantumly dressed operator with the above nontrivial conformal dimension. 

On the dual gravity side, the relevant deformation generally modifies both the inner geometry and the minimal surface area.
In \cite{Hung:2011ta}, only the first order deformation of the minimal surface has been taken into account. As will be seen, 
it trivially vanishes due to the symmetry of the original action. However, higher order deformations still survive and nontrivially contribute the entanglement entropy. In this work, we will investigate the entanglement entropy generated by higher order minimal surface deformations. Interestingly, we find that such higher order deformations can yield an additional logarithmic correction. In a disk-shaped region for $d=3$ and $4$, we explicitly calculate the entanglement entropy and show that the second order minimal surface deformation leads to an additional logarithmic term. It together with the result obtained by the metric deformation determines the coefficient of the logarithmic correction exactly because there is no more logarithmic term caused by higher order metric and minimal surface deformations.

The rest of this paper is organized as follows: In Sec. 2, we summarize higher dimensional holographic entanglement entropies and the conformal perturbation theory deformed by a relevant operator. In Sec. 3, we summarize a geometry deformed by a massive scalar field and a general logarithmic term caused by the metric deformation. In addition, we account for the possibility of an additional logarithmic term caused by the minimal surface deformation.
 In Sec. 4, we explicitly calculate the entanglement entropy of the deformed theory and shows that there really exists an
 additional logarithmic term caused by the second order minimal surface deformation for $d=4$ and $3$ examples.
 We finish our work with some concluding remarks in Sec. 5.


\section{Relevant perturbation in CFT}

The entanglement entropy is an important concept to understand  quantum aspects of a QFT. It measures the entanglement between quantum states.  In a two-dimensional CFT, especially, its analytic form has been known due to the conformal symmetry and modular invariance. Interestingly, this result has been reproduced from a gravity theory defined on the AdS$_3$ space and further generalized to higher dimensional cases \cite{Ryu:2006bv,Ryu:2006ef}. We begin with summarizing those results.

In a $d$-dimensional CFT, when a subsystem lies in a thin strip with a width $l$, the holographic entanglement entropy is given by the area of a minimal surface extended in AdS$_{d+1}$
\be			\la{res:stritpEE}
A = \fr{2 R^{d-1}}{d-2} \ls \fr{L}{\e} \rs^{d-2} - \fr{2^{d-1} \pi^{(d-1)/2} R^{d-1}}{d-2}
\lb \fr{\G \ls \fr{d}{2(d-1)}\rs }{\G \ls \fr{1}{2(d-1)}\rs }\rb^{d -1} \ls \fr{L}{l} \rs^{d-2} ,
\ee
where $L$ and $\e$ indicate a size of the total system and a UV cutoff respectively. The AdS radius is denoted by $R$. From now on, we set $R=1$ for simplicity. This result expresses the entanglement entropy of vacuum states. Above the first term shows the area law of the entanglement entropy. In general, the entanglement entropy of a thin strip has no logarithmic divergence except the $d=2$ case. When a subsystem resides in a disk instead of a strip, the entanglement entropy usually depends on the dimension of the space on which the CFT is defined. When $d$ is odd it becomes \cite{Ryu:2006bv}
\be		\la{res:odddiskEE}
A= \O_{d-2} \lb \fr{1}{d-2} \ls \fr{l}{\e}\rs^{d-2}  + F + {\cal O} \ls \fr{\e}{l} \rs \rb ,
\ee
while it for $d=$even gives rise to
\be		\la{res:evendiskEE}
A= \O_{d-2} \lb \fr{1}{d-2} \ls \fr{l}{\e}\rs^{d-2}  +  a' \ \log \ls \fr{l}{\e} \rs + {\cal O} (1) \rb .
\ee
Only for even $d$, a logarithmic term with $a' = (-)^{d/2 -1} \fr{(d-3)!!}{(d-2)!!}$ appears. This term is universal in that it is independent  of the regularization scheme. As shown in the AdS$_3$ example \cite{Ryu:2006bv,Ryu:2006ef}, $a'$ is related to the central charge of the dual CFT. In a higher dimensional CFT, $a'$ is  related to an A-type central charge. As a consequence, the logarithmic term related to the central charge crucially depends on the dimension and shape of the entangling surface. 

Recently, it has been discussed that a relevant operator with the specific conformal dimension can provide an additional logarithmic term \cite{Hertzberg:2010uv}, which has been checked in a free CFT \cite{Huerta:2011qi,Lewkowycz:2012qr,Rosenhaus:2014woa,Rosenhaus:2014ula,Rosenhaus:2014zza}. In a conventional QFT defined on a Euclidean space, the action is given by a functional of the metric and fields. If the system resides in the vacuum state $\left| 0 \ket$, the reduced density matrix of a subsystem $A$ is represented as the trace of the density matrix over its complement denoted by $\bar{A}$
\be 		
\r_0 = \Tr_{\bar{A}} \ \left| 0 \ket \bra 0 \right| \equiv e^{-K_0}  ,
\ee
where $K_0$ means the modular Hamiltonian. In general, the modular Hamiltonian is unknown except several cases, planar and spherical entangling surfaces. For a planar entangling surface embedded in a flat space, the modular Hamiltonian is proportional to the Rindler Hamiltonian \cite{Rosenhaus:2014woa,Rosenhaus:2014ula,Rosenhaus:2014zza}
\be
K_0 = - 2 \pi \int_{\S} d^{d-2} x\int_0^{\infty} d x_1 \ x_1 \ T_{2 2} ,
\ee
where $\S$ is the entangling surface. Here coordinates, $x_{\m}=\lc x_a , y_i\rc$, indicate transverse and longitudinal directions along the plane surrounded by the entangling surface. Above $x_1$ and $x_2$ correspond to orthogonal coordinates. In this case, the entanglement entropy is expressed as the Von-Neumann entropy
\be   	\la{def:entent}
S_0 = - \Tr \ls  \r_0 \log \r_0 \rs = \bra 0 \left| K_0 \right| 0 \ket ,
\ee
where the last term indicates a Euclidean path integral over the entire manifold with insertion of $K_0$ \cite{Casini:2011kv,Rosenhaus:2014woa,Rosenhaus:2014ula,Rosenhaus:2014zza}. 

Now, let us deform the CFT with a relevant operator ${\cal O}$. Then, the deformed modular Hamiltonian can be written as
\be		\la{def:fermionmass}
K = K_0 + \l {\cal O} \equiv K_0 + \l  \int d^4 x \  {\cal O} (x) ,
\ee
where $K_0$ denotes the modular Hamiltonian of the undeformed CFT. Expanding the deformed entanglement entropy, $S =  \Tr \ls e^{- K } K \rs$, in the small coupling limit ($\l \ll 1$), its first order variation with respect to the coupling constant leads to 
\be
\lp \fr{\pa S}{\pa \l} \right|_{\l=0} = - \bra K {\cal O} \ket_{\l=0} + \bra \fr{\pa K}{\pa \l} \ket_{\l=0}
= - \bra K_0 {\cal O} \ket + \bra {\cal O} \ket ,
\ee
where $\bra \cdots \ket$ indicates $\Tr \ls e^{-K_0} \cdots \rs$. In order to obtain the last equality we used the fact that the undeformed theory is independent of $\l$. Note that the last term automatically vanishes, $\bra {\cal O} \ket=0$,  due to the normalization, $\Tr e^{- K_0}=\Tr e^{- K}= 1$.  The variation of the entanglement entropy at second order yields
\be
\lp \fr{\pa^2 S}{\pa \l^2} \right|_{\l=0}  = \bra K {\cal O} {\cal O} \ket_{\l=0} - \bra  \fr{\pa K}{\pa \l} {\cal O}  \ket_{\l=0} 
= \bra K_0 {\cal O} {\cal O}  \ket - \bra {\cal O} {\cal O} \ket .
\ee 
Combining these results, the change of the entanglement entropy up to second order can be written as the sum of correlation functions between the deformation operator and the undeformed modular Hamiltonian 
\be		\la{eq:relevantdef}
\d S = - \bra K_0 {\cal O}  \ket  \l + 
\fr{1}{2} \ls \bra K_0 {\cal O} {\cal O} \ket - \bra {\cal O} {\cal O} \ket \fr{}{} \rs    \l ^2
+ \cdots .
\ee
The correlator $\bra T_{\m\n} {\cal O} \ket $ vanishes for a CFT,  so does $\bra K_0 {\cal O} \ket$ because $K_0 \sim T_{\m\n}$ \cite{Rosenhaus:2014zza}. The first nonvanishing contribution consequently appears at $\l^2$ order. When CFT correlation functions are given by \cite{Rosenhaus:2014zza,Osborn:1993cr,Erdmenger:1996yc}
\bea
\bra {\cal O} (x_2) {\cal O} (x_3) \ket &=&  \fr{N}{(x_2 - x_3)^{2 \D}} , \la{res:twofunction} \\
\bra T_{\m\n} (x_1)  {\cal O} (x_2) {\cal O} (x_3) \ket &=& \fr{1}{x_{12}^d x_{23}^{2\D -d}
x_{31}^d}  t_{\m\n} (\hat{X}_{23}) ,
\eea 
where
\be
t_{\m\n} (\hat{X}_{23}) = \fr{d \D N}{(d-1) S_d} \ls \hat{X}_{\m} \hat{X}_{\n} - \fr{1}{d} \d_{\m\n} 
\rs \quad {\rm and} \quad X_{23} = \fr{x_{21}}{x_{21}^2} -  \fr{x_{31}}{x_{31}^2} ,
\ee
with
\be
\hat{X}_{\m} = \fr{X_{\m}}{\sqrt{X^2}} \ , \quad  x_{ij} = x_{i} - x_{j}  ,
\ee
their integrals give rise to \cite{Rosenhaus:2014zza}
\bea
\bra {\cal O} {\cal O} \ket 
&\equiv& \int d^d x \int d^d y \ \bra {\cal O} (x)  {\cal O} (y) \ket \nn
&=& \fr{N \pi^{d/2+1}}{\D-(d-1)/2} \fr{\G (\D-d/2) A_{\S}}{\G(\D)}
\int_{\d}^l dx_1 \ x_1^{d - 2 \D +1} ,
\eea
and 
\bea
\bra K_0 {\cal O} {\cal O} \ket &=& (2 \pi)^2 \int d^{d-2} y \int_0^{\infty} d x_1 x_1
\int d^{d-2} \bar{y} \int^0_{-\infty} d \bar{x}_1 \bar {x}_1 \int d^d z
\bra T_{22} (x) {\cal O} (\bar{x}) {\cal O} (z)\ket \nn
&=& \fr{4 d \D N \pi^{d+1}}{d (d-1) (d-2\D) (d-2 \D -1) S_d}  \fr{\G (\D-d/2) A_{\S}}{\G(\D) \G(d/2)}
\int_{\d}^l dx_1 x_1^{d-2 \D +1} ,
\eea
where $l = \l^{1/(\D-d)}$ is introduced as an IR cutoff.

Now, consider a free massless fermion theory as the undeformed CFT and deform it by a fermion mass term, $\l = m$ and ${\cal O}= \bar{\ps} \ps$. In this case, the deformation operator corresponds to the relevant operator with the conformal dimension $\D=3$. From the fermion propagator of the free massless fermion theory
\be
\bra \bar{\ps} (x) \ps (0) \ket = \fr{1}{S_4} \fr{\g_{\m} x^{\m}}{x^4} ,
\ee
where $\g_{\m}$ is a Euclidean gamma matrix and $S_4$ denotes a solid angle, the two-point correlation function of ${\cal O}$ reads
\be			\la{res:twopointO}
\bra {\cal O} (x) {\cal O} (0) \ket = \fr{4}{S_4^2} \fr{1}{x^{6}} .
\ee
Note that, if one takes into account an interacting fermion theory as the undeformed CFT, the conformal dimension of a fermion field and the stress tensor may be modified due to the interaction and quantum corrections. Comparing \eq{res:twopointO} with \eq{res:twofunction}, the normalization constant is given by
$N = \fr{4}{S_4^2}$. 
In a small mass limit, this relevant deformation causes a small change of the entanglement entropy according to \eq{eq:relevantdef}. Using the above normalization constant, the small change of the entanglement entropy leads to the following logarithmic correction at second order \cite{Huerta:2011qi,Lewkowycz:2012qr,Rosenhaus:2014zza}
\be		\la{res:logcorrection}
\d S = \fr{1}{12 \pi}  \ m^2 A_{\S} \log \ls m \e \rs .
\ee
In sum, the additional logarithmic correction in a $d$-dimensional CFT occurs at $\l^2$ order when the relevant operator has the conformal dimension $\D = \fr{d+2}{2}$. It would be interesting to ask whether the operator with the same conformal dimension can generate a similar logarithmic correction in a strongly interacting CFT. In the next section, we will look into this issue by using the AdS/CFT correspondence.

\section{General logarithmic correction in the strong coupling regime}

In a $d$-dimensional free CFT, a scalar operator composed of $N_B$ bosons and $2 N_F$ fermions has the following conformal dimension
\be
\D_{free} = \fr{d-2}{2} N_B + \ls d -1\rs N_F ,
\ee 
where the number of fermions should be even for the Lorentz invariance. In a free CFT, we can easily see that
 there is no operator with $\D=\fr{(2n-1)d+2}{2 n}$ for $n \ge 2$ because $\D_{free}$ is always bigger than $\D$. However, this is not true in an interacting theory because quantum corrections  alters the conformal dimension through the anomalous dimension. In the large conformal dimension limit,  anomalous dimensions of the scalar operators have been widely investigated in \cite{Minahan:2002ve}-\cite{Lee:2011fe}. Furthermore, the anomalous dimension of the Konishi operator with a low conformal dimension has been studied in \cite{Anselmi:1996dd,Bajnok:2008bm,Ahn:2010yv,Roiban:2009aa}. Due to these quantum effects, scalar operators with $\D=\fr{(2n-1)d+2}{2 n}$ may exist in an interacting CFT. 

Let us consider a CFT deformed by a scalar operators, ${\cal O}$, with $\D=\fr{(2n-1)d+2}{2 n}$ in the strong coupling limit. According to the AdS/CFT correspondence, it can be dual to the following gravity theory
\be \la{action:chiral}
S = \int d^{d+1} x \sqrt{-G} \lb \frac{1}{2 \k^2} \ls {\cal R} - 2 \L \rs
- \frac{1}{2}  \ls \pa \ph \rs^2  -  \frac{1}{2} m_{\ph}^2 \ph^2  \rb .
\ee
If we denote a source as $\l$, it should scale $\l \to \O^{-\D_s} \l$ with $\D_s = \fr{d-2}{2n}$ under $x^{\m} \to \O x^{\m}$. This scaling behavior is encoded into the asymptotic solution of a dual bulk field with the following mass 
\be
m_{\ph}^2 = - \fr{(d-2) \lb (2n -1)d + 2 \rb}{4 n^2} ,
\ee
which is always bigger than the Breitenlohner-Freedman bound, $- d^2/4$, for $d>2$. Therefore, the deformation we consider is stable in the AdS geometry.  Since the relevant deformation does not significantly change the theory in the asymptotic region,  we can take the following metric ansatz 
\be
ds^2 = \fr{1}{z^2}  \ls f(z) \et_{\m\n} dx^{\m} dx^{\n}  + dz^2 \rs ,
\ee
where the deviation from AdS is encoded into $f(z)$. The perturbative solution of $\ph$ in the asymptotic region
is expanded into
\be		\la{sol:gpersol}
\ph =  \l z^{\D_s}  \ls 1 + \cdots +  c z^{\D-\D_s} \log z   + \cdots \rs + \bra {\cal O} \ket  z^{\D}  \ls 1 + \cdots \rs ,
\ee 
where $\l$ and $\bra {\cal O} \ket $ are two integration constants. Note that since the deformation we consider is proportional to $\ph^2$, the deformed theory should be invariant
under $\ph \to - \ph$ ($\l \to - \l$ and $\bra {\cal O} \ket \to - \bra {\cal O} \ket$). 
When $\D/\D_s$ is an integer, two independent
solutions of the second order differential equation become degenerate at $z^{\D}$ order. Therefore, the correct perturbative solution requires an additional logarithmic term at $z^{\D}$ order. For $d=3$ and $d=4$, since $\D/\D_s$  is always an integer, the perturbative solution of the deformed AdS geometry should have a logarithmic term. 
This is the reason why we include $c  z^{\D} \log z$ in \eq{sol:gpersol} where $c$ is again fixed by $\l$ and $\bra {\cal O} \ket $. If $\D/\D_s$ is not an integer, there is no such logarithmic term so that $c$ must be zero. 
Then, its gravitational backreaction in the asymptotic region is given by 
\be   \la{res:gmetsoll}
f(z)  = 1 + d_1 \l^2 z^{2 \D_s }+ \cdots 
+ d_n \l^{2n} z^{2n \D_s} + \cdots + \ls d_{d-\D_s} + d_{\D}' \log z \rs z^{d} + \cdots ,
\ee 
where all $d_n$'s are again uniquely determined in terms of two integral constants. The logarithmic term in the metric appears due to the combination of $ \l z^{\D_s}  $ and $c \l z^{\D} \log z$ 
in $\ph$. Note that the above metric involves a $z^{d-2}$ term at $n$-th order. This metric deformation leads to a generalized logarithmic correction \cite{Hung:2011ta}
\be   	\la{res:newadditional}
\d S \sim \l^{2n} A_{\S} \log \ls \l^{2n/(d-2)} \e\rs .
\ee

Now, let us take into account the entanglement entropy in a disk-shaped region. To obtain it for $ \l l^{\D_s } \ll 1$, we need to expand the minimal
surface near the known solution, $z_0 = \sqrt{l^2 - \r^2}$,
\be
z (\r) = z_0 + \sum_{i=1} \ls \l l^{\D_s } \rs^i  z_i  .
\ee
 In \cite{Hung:2011ta}, authors have considered the first order minimal surface deformation, which does not contribute to the entanglement entropy because $z_1$ is not invariant under $\l \to - \l$. In general, $z_i$ with $i=odd$ automatically vanishes due to the invariance under $\l \to - \l$. When $i$ is $even$, there exist nontrivial minimal surface deformations contributing to the entanglement entropy. Intriguingly, the $2n$-th order minimal surface deformation involves 
 $\l^{2n} l^{2n \D_s} = \l^{2n} l^{d-2}$ which can be rewritten as $\l^{2n} A_{\S} $ after multiplying the solid angle of the entangling surface, $\O_{d-2}$. This implies that the $2n$-th order minimal surface deformation can gives rise to
the same coupling dependence appearing in the logarithmic term in \eq{res:newadditional}. In other words, there is the possibility to obtain a new logarithmic correction caused by the minimal surface deformation. In the next section, we will
investigate such a new logarithmic correction generated by the minimal surface deformation.
For $n>1$, it may be possible to obtain more general logarithmic terms caused by the composition of the metric and minimal surface deformations. We leave this issue as a future work.

\section{Relevant deformation in the strong coupling regime}

In Sec. 2, we showed that a relevant operator with $\D=(d+2)/2$ provides an additional logarithmic term in a free fermion theory. In order to understand the deformation of a strongly interacting CFT, we need to take into account the dual gravity theory according to the AdS/CFT correspondence. In the AdS/CFT context, the relevant deformation operator can be realized by introducing an appropriate massive scalar field to the AdS geometry. In this case, the gravitational backreaction of the scalar field modifies the inside geometry which is associated with the IR deformation of the dual CFT. If the mass of the scalar field is given by $m_{\ph}$, the conformal dimension of the dual operator is determined from
\be
m_{\ph}^2 = \D \ls \D - d \rs .
\ee
When we take $m_{\ph}^2=-3$ for $d=4$, the dual operator represents a scalar operator with $\D = 3$. If a fermion field, $q$, in a four-dimensional CFT has no anomalous dimension, a possible candidate of the operator with $\D = 3$ may become ${\cal O}_q=\bar{q} q$. Its vacuum expectation value (vev) is called the chiral condensate, $\s=\bra {\cal O}_q\ket$, which has been widely studied in the holographic QCD model to understand the chiral symmetry breaking effect \cite{Erlich:2005qh,Da Rold:2005zs,Karch:2006pv,Sakai:2004cn,Sakai:2005yt,Lee:2010dh,Park:2011ab}.  For $d=3$, on the other hand, we need to take $m_{\ph}^2=-5/4$  for generating a logarithmic correction because it corresponds to a scalar operator with $\D=5/2$. 

Assuming that $\ph$ in \eq{action:chiral} depends only on the radial coordinate, then the relevant deformation operator we consider does not affect the asymptotic geometry. In this case, the scalar field for $d=4$ has the following perturbative solution \cite{Lee:2010dh,Park:2011ab,Lee:2013oya,Lee:2015rxa}
\be
\ph (z) = m_q z \ls 1+ \cdots \rs + \s  z^3  \ls 1+ \cdots \rs .
\ee
Here, $m_q$ behaves as a source of ${\cal O}_q$ and corresponds to the current quark mass. This fact becomes clear when regarding the on-shell action of the scalar field
\be
\int d^4 x \ m_q  \bar{q} q ,
\ee
which is the exact same as the relevant deformation given in \eq{def:fermionmass}. The gravitational backreaction of this scalar field does not break the Poincare symmetry of the boundary space because it depends only on the radial coordinate. Therefore, the most general ansatz including the gravitational backreaction is given by the following form  
\be
ds^2 = \frac{1}{z^2} \lb f(z) \ls - dt^2 + d \vec{x}^2 \rs  + dz^2 \frac{}{} \rb .
\ee
Note that since the mass deformation is proportional to $\ph^2$, the deformed theory should be invariant
under $\ph \to - \ph$. In other words, the theory we considered should be invariant under 
$m_q \to - m_q$ and $\s \to - \s$. 

Requiring that the above metric reduces to AdS in the asymptotic region, $f(z)$ should approach to $1$ at the boundary $z=0$. Together with this constraint, the perturbative solution satisfying the equations of motion 
\bea    \la{eq:chiral}
0 &=& \k^2 f \ls 3 \ph^2 - z^2 \ph'^2 \rs - 6 z \ls z f'' -3 f' \rs  ,\nn
0 &=&  \frac{6 f'^2}{f^2} - \frac{3 \k^2 \ph^2}{z^2} - \frac{24 f'}{z f}
- \k^2 \ph'^2  , \nn
0 &=&  2 z^2 f' \ph' + f \ls 3 \ph + z ( z \ph'' -3 \ph' ) \rs  ,
\eea
is given by \cite{Lee:2010dh,Park:2011ab}
\bea    \la{fmqnzero}
f(z)&=&1 - \frac{\k^2}{6}   m_q^2 z^2 +
 \frac{1}{36} \ls \frac{}{} \k^4 m_q^4 - 9 \k^2 m_q \s - 3 \k^4 m_q^4 \log z \rs z^4 
 + {\cal O} (z^6)  , \nn
\ph (z) &=& m_q z + \ls \s + \frac{1}{3} \k^2 m_q^3 \log z \rs z^3  +  {\cal O} (z^5) .
\eea

From this perturbative solution one can find an exact numerical solution, which generally yields a singularity at the center, $z=\infty$. This fact indicates that this solution is IR incomplete.  However, since the logarithmic term we are interested in is insensitive to IR details \cite{Lewkowycz:2012qr,Hung:2011ta}, the above solution is sufficient to study the logarithmic correction. Nevertheless, it still remains as an interesting question to figure out IR physics. One way to get rid of the IR incompleteness is to introduce higher order potential terms like $\l_n \ph^n$ ($n>2$).  In \cite{Hung:2011ta}, the entanglement entropy for $n=3$ has been investigated. In the UV region,  higher order potential terms usually generate higher order corrections and  do not contribute to the logarithmic correction. On the contrary, they play a crucial role in the IR regime. In the present paper, we do not discuss on the IR completion although it deserves to study further its effect in the IR regime.

\subsection{Entanglement entropy in a strip}

Let's consider the entanglement entropy in a thin strip. Suppose that a CFT is defined in a regularized spatial volume
\be
- \fr{L}{2} \le x_1,x_2 , x_3 \le \fr{L}{2} ,
\ee
and that the total system is divided into a subsystem and its complement. When the strip is represented as
\be
- \fr{l}{2} \le x_1 \equiv x \le \fr{l}{2}  \quad {\rm and}  \quad - \fr{L}{2} \le x_2 , x_3 \le \fr{L}{2} ,
\ee
the holographic entanglement entropy can be measured by the area of a minimal surface extended in the AdS space \cite{Ryu:2006bv,Ryu:2006ef}. In this case, the boundary of the minimal surface has to coincide with the entangling surface of the strip. Then, the minimal surface is governed by the following action 
\be
A = \int d^3 x \ \sqrt{h} .
\ee
Here $h_{\m\n}$ indicates an induced metric on the minimal surface
\be
ds_{in}^2 = \fr{1}{z^2} \lb \ls f(z) + z'^2 \rs  
dx^2 + f(z)  dx_2^2 + f(z)  dx_3^2 \fr{}{} \rb  ,
\ee
where the prime means a derivative with respect to $x$. Due to the symmetry under $x \to - x$, the minimal surface has a turning point at $x=0$. If we denote this turning point  as $z_*$, the following relation is automatically satisfied at the turning point 
\be
\lp z' \right|_{z=z_*} = 0 .
\ee
From this fact, the conserved quantity leads to the following first order differential equation which relates the turning point to the width of the strip, 
\be		\la{eq:stripwidth}
z' = \fr{z_*^3 \ f(z)^2}{z^3 \ f(z_*)^{3/2}}  \sqrt{1 - \fr{z^6 \ f(z_*)^3}{z_*^6 \ f(z)^3}} .
\ee
Substituting this relation back into the action, the area of the minimal surface is rewritten as the integral form
\be
A = 2 L^2 \int_{\e}^{z_*} dz \fr{f(z)}{z^3} \fr{1}{\sqrt{1 - \fr{z^6 \ f(z_*)^3}{z_*^6 \ f(z)^3}}} ,
\ee
where $\e$ is introduced as a UV cutoff and the range of $z$ is restricted to $\e \le z \le z_*$.

Since the logarithmic correction appears in the UV region as mentioned before, it is sufficient to consider only the limit with $m_q z_* \ll 1$ and $\s z_*^3 \ll 1$, where the minimal surface is extended only in the UV region. Taking this limit implies that we take into account a very thin strip with $m_q l \ll 1$. The existence of small parameters makes the perturbative calculation possible. From \eq{eq:stripwidth} the strip width is determined in terms of $z_*$
\bea		\la{res:widthstripz0}
l &=&  2 \int_0^{z_*} dz \ls \fr{ z^3}{\sqrt{z_*^6 - z^6} } 
- \fr{\k^2 m_q^2  \ls z^9 + z^7 z_*^2 + z^5 z_*^4 - 3 z^3 z_*^6 \rs }{
 12 (z^4 + z^2 z_*^2 + z_*^4) \sqrt{z_*^6 -z^6}}   + \cdots \rs  \nn
 &=& \fr{ \sqrt{\pi} \G \ls \fr{5}{3} \rs}{2 \G \ls \fr{7}{6} \rs}  z_* 
 - 0.039 \k^2 m_q^2 \ z_*^3 + \cdots .
\eea
Rewriting $z_*$ in terms of $l$ gives rise to
\be
z_* =  \fr{2 \G \ls \fr{7}{6} \rs}{ \sqrt{\pi} \G \ls \fr{5}{3} \rs} l 
- 0.071 \k^2 m_q^2  l^3 + {\cal O} \ls m_q^4  l^5 \rs .
\ee
Similarly, the area of the minimal surface can be expanded into 
\bea
A 
 &=&  L^2 \int_{\e}^{z_*} dz \fr{ 2  z_*^3}{z^3 \sqrt{z_*^6 - z^6}}
-  L^2 m_q^2 \k^2 \int_{\e}^{z_*} dz 
\frac{z_*^3 \ls 5 z^4+2 z^2 \text{z_*}^2+2 \text{z_*}^4 \rs}{6 z
   \left(z^4+z^2
   \text{z_*}^2+\text{z_*}^4\right) \sqrt{{\text{z_*}^6}-{z^6}} }  + \cdots \nn
 &=&  \frac{L^2}{\epsilon ^2}
 - \fr{ L^2  m_q^2 \k^2}{3} \log \fr{z_*}{\e} 
  -\frac{\sqrt{\pi } L^2 \Gamma \left(\frac{5}{3}\right)}{4 z_0^2 \Gamma \left(\frac{7}{6}\right)}
 -0.172 L^2 m_q^2 \k^2
 + {\cal O} (m_q^4 z_*^4) .
\eea  

Finally, the resulting entanglement entropy reads in terms of $l$
\bea
S \equiv \fr{2 \pi A}{\k^2}
= \ \frac{\pi  A_{\Sigma}}{\kappa ^2 \epsilon ^2}
+\frac{\pi  m_q^2 A_{\Sigma }  }{3} \log \left( m_q \epsilon  \right)
+ {\cal O} (1) ,
\eea
where $A_{\Sigma} = 2 L^2$ means the area of the entangling surface. The first divergent term comes from the undeformed CFT, while the second corresponds to the first correction caused by the relevant deformation. Similar to the free fermion case, the above result shows that the relevant operator with the specific conformal dimension generates a similiar logarithmic correction even in a strongly interacting CFT
\be		\la{res:untermstrip4}
\d S = \fr{\pi}{3} m_q^2 A_{\S} \log \ls m_q \e \rs  .
\ee
This additional correction was called a universal logarithmic term in that it is always proportional to the entangling surface area unlike the one associated with the central charge in \eq{res:evendiskEE} \cite{Hertzberg:2010uv,Huerta:2011qi,Lewkowycz:2012qr,Rosenhaus:2014woa,Rosenhaus:2014ula,Rosenhaus:2014zza}. Note that this result is coincident with the result in \cite{Hung:2011ta} and parameters we used are related to those in \cite{Hung:2011ta}
by $\d = m_q \e^2$ and $m_q = \l \m$.

\subsection{Entanglement entropy on a disk}

In general, the logarithmic term related to the A-type central charge significantly depends on the dimension and shape of the entangling surface. So we can ask how the additional logarithmic correction caused by the specific operator relies on properties of the entangling surface. In order to understand such dependencies, in this section, we will investigate the entanglement entropy contained in a disk instead of a strip.

For regarding a subsystem defined in a disk, it is more convenient to take the following metric form due to the rotational symmetry
\be
ds^2 = \frac{R^2}{z^2} \lb f(z) \ls - dt^2 + d \r^2 + \r^2  d \O_2^2 \rs  + dz^2 \frac{}{} \rb ,
\ee
where $ d \O_2^2$ denotes a metric of a two-dimensional unit sphere. Then, a disk (more precisely, three-dimensional ball) can be represented as
\be
0 \le \r \le l \quad {\rm and} \quad z=z(\r)  .
\ee
Using this parametrization, the action describing a minimal surface is reduced to
\be		\la{act:5ddisk}
A = \O_2 \int_{0}^{l} d \r  \ \fr{\r^2 f(z)}{z^3} \sqrt{ f(z) + z'^2} .
\ee
Since this action explicitly depends on $\r$, there is no conserved quantity under the translation in the $\r$ direction. Thus, the conservation law cannot be applied directly. Instead, we can use the known solution, $z_0 (\r)$, in the pure AdS geometry \cite{Ryu:2006bv,Ryu:2006ef,Kim:2014yca,Kim:2014qpa}. Using dimensionless small parameters, $m_q l$ and $\s l^3$, the deformed minimal surface near the known solution can be expanded into
\be     \la{exp:disk4ansatz}
z (\r)= z_0 (\r) + m_q l \ z_2 (\r)  + m_q^2 l^2 \ z_2 (\r) + \cdots,
\ee
where terms involving $\s l^3$ appear at higher orders.

At leading order, the action reduces to
\be		
A_0 = \O_2 \int_{0}^{l} d \r  \ \fr{\r^2}{z_0^3} \sqrt{1+z_0'^2} ,
\ee
and describes the minimal surface area of the undeformed theory. Its equation of motion reads
\be   \la{eq:zerothorder}
0 =  3 r \left( z_0'^2+1 \right)+ z_0 \left( r z_0''+2 z_0'^3+2 z_0' \right) .
\ee
It has been known that this equation allows a solution corresponding to a geodesic of a particle in the AdS space \cite{Ryu:2006bv,Ryu:2006ef}, which is given by
\be		\la{res:leadingsol}
z_0 =\sqrt{ l^2 - \r^2} .
\ee 
This solution satisfies the following two boundary conditions, $z_0(l)  = 0$ and $z_0' (0) = 0$. The former indicates that the boundary of the minimal surface lies at $z=0$, while the latter is required to make the minimal surface smooth at the turning point, $\r=0$. Because of the fixed entangling surface and the smoothness of the minimal surface, the deformed minimal surface should also satisfy the same boundary conditions. This fact plays an important role in determining all higher order corrections uniquely.

At $m_q l$ order, the action is given by
\be
A_1 = \text{m_q} l \ \O_2 \int_{0}^{l} d \r  \ 
\frac{ r^2 \left( z_0 z_0' z_1'  -3   z_0'^2 z_1 -3 z_1 \right)}{z_0^4
   \sqrt{ 1+ z_0'^2}} .
\ee
At this level, $z_1$ is not fixed because there is no kinetic term. To determine it
we should go to the next order.
The action at $m_q^2 l^2$ order consists of three parts
\be
A_2 = {A}_{20} + {A}_{21} + {A}_{22} ,
\ee
\bea
{A}_{20} &=& 
-  \text{m_q^2} l^2 \ \O_2  \int_{0}^{l} d \r  \
\frac{ \kappa ^2 \rho ^2 \left(3 l^2 - \rho ^2 \right) }{12 l^3 \left(l^2-\rho ^2\right)} , 
\la{eq:2nd1} \\
{A}_{21} &=& 
\text{m_q^2} l^2 \ \O_2  \int_{0}^{l} d \r  \ 
\frac{\rho ^2 \left(12 l^4 z_1^2+6 l^2 \rho  \left(l^2-\rho
   ^2\right) z_1 z_1'+\left(l^2-\rho
   ^2\right)^3 z_1'^2\right)}{2 l^3 \left(l^2-\rho
   ^2\right)^3},  \la{eq:2nd2}\\
{A}_{22} &=& 
- \text{m_q^2} l^2 \ \O_2  \int_{0}^{l} d \r  \ 
\frac{\rho ^2 \left(l^2 \rho  \left(l^2 - \rho ^2 \right) z_2' + 3 l^4
   z_2 \right)}{l^3 \left(l^2-\rho ^2\right)^{5/2}}  , \la{eq:2nd3}
\eea
where the leading solution in \eq{res:leadingsol} is used.
Above the first represents the contribution from the metric deformation, while
the remaining are related to the first and second order deformations of the minimal surface. 
From  \eq{eq:2nd2}, the equation of motion for the first order minimal surface deformation reads
\be
0 = \left(l^2-\rho ^2\right)^2 \left(\rho  z_1''  + 2 z_1'\right)-3 l^2 \rho  z_1,
\ee
which allows the following solution
\be
z_1= \frac{c_1 (l - \rho)^2}{\rho  \sqrt{l^2 - \rho^2 }}+\frac{c_2}{\sqrt{l^2 - \rho^2 }} .
\ee
To satisfy $z_1 = 0$ at $\r=0$, $c_2$ must vanish. In addition, $c_1$ should also be zero because $z_1'$ diverges at $\r=0$. Thus,
there is no nontrivial $z_1$ satisfying the above two boundary conditions which is consistent with the previous symmetry 
description. As mentioned before, the theory we consider should be invariant under $\ph \to - \ph$. Thus, the terms with an odd power in \eq{exp:disk4ansatz} should automatically vanish. As a result, there is no entanglement entropy change caused by the first order minimal surface deformation which has already been argued in \cite{Hung:2011ta}. This automatically implies  $A_1=A_{21}=0$. However, the terms with an even power can still survive and give rise to a nontrivial contribution to the entanglement entropy. 

Now, let us consider the second order minimal surface deformation described by $z_2$.  
At $m_q^2 l^2$ order, the remaining terms are
\be
A_2 = {A}_{20} + {A}_{22}  .
\ee
Here, the first represents the contribution from the metric deformation studied in \cite{Hung:2011ta}, while the second is related to the second order deformation of the minimal surface. Since the second order minimal surface deformation occurs at the same level as the metric deformation causing the logarithmic term, it would be interesting to ask whether the second order minimal surface deformation can provide an additional logarithmic correction. To do so, let us first find the solution of $z_2$. Note that $z_2$ is not fixed from ${A}_{22}$ because it does not contain a kinetic term.  
Thus, we need to look into the $m_q^4 l^4$ order terms for obtaining the equation of motion of $z_2$.  At quartic order the equation of motion reads
\be
0= 3 l^2 \rho  \left(l^2-\rho ^2\right)^2 z_2''+6 l^2
   \left(l^2-\rho ^2\right)^2 z_2' -9 l^4 \rho  z_1+\kappa ^2 \rho ^3 \left(l^2-\rho
   ^2\right)^{3/2} .
\ee
An exact solution of this inhomogeneous differential equation is given by\footnote{We thank N. Kim for noting that there exists an analytic solution.}
\bea
z_2 (\r) &=& \frac{c_3 (l-\rho )^{3/2}}{\rho  \sqrt{l+\rho }}+\frac{c_4}{\sqrt{l^2-\rho^2 }} \nn
&& +\frac{\kappa ^2 \left(12 l^3 (l-\rho )^2
   \tanh ^{-1}\left(\frac{\rho }{l}\right)-\rho  \left(-24 l^4 \log (l+\rho )+12 l^4+4 l^2 \rho ^2+\rho ^4\right)\right)}{36 l^2 \rho \sqrt{l^2-\rho ^2}} .
\eea
In order for this solution to describe a well-defined minimal surface, it should satisfy previous two boundary conditions. The first one, $z_2 (l) = 0$, determines $c_4$ as 
\be
c_4 =\frac{\lb 17 \sqrt{2}-24 \sqrt{2} \log (2 l) \rb  \ \kappa ^2 l^{2}}{36 \sqrt{2}} ,
\ee
and the other, $z'(0) = 0$, fixes $c_3$ to be zero. 
Near the asymptotic boundary, the first correction of the minimal surface deformation reads
\be
z_2 \approx 
\frac{ \k^2  (l-\rho )^{3/2}  \left(-13 \sqrt{2} +6 \sqrt{2}  \log  2 -6 \sqrt{2}  \log (l-\rho ) +6 \sqrt{2}   \log  l  \right)}{72 \sqrt{l}} .
\ee

Using these solutions, the fermion mass leads to the deformation of the minimal surface area, $A = A_0 + {A}_{20} + {A}_{22} $, up to $m_q^2 l^2$ order. Rewriting it in terms of $x= l -\r$, we reach to
\bea
A_0 &=& \Omega _2 \int_{x_*}^l dx \ \frac{ l-x}{4 x^2} + {\cal O} (1) , \nn
{A}_{20} &=& -  m_q^2 l^2 \Omega _2  \int_{x_*}^l dx \ \frac{\kappa ^2}{12 x}
 + {\cal O} (1) , \nn
{A}_{22} &=& - m_q^2 l^2 \Omega _2   \int_{x_*}^l dx \ \frac{ \k^2  }{24 x} + {\cal O} (1)  ,
\eea 
where the lower bound, $x_*$, is introduced to denote a UV cutoff in the $x$-direction which is associated with the UV cutoff in the $z$-direction, $\e$. To understand their relation, let us recall that the full solution can be approximated by $z \approx \sqrt{ 2 l x}$. From this, we can derive the following relation 
\be
x_* = \fr{\e^2}{2 l} + {\cal O}( \e^4)  ,
\ee
where the effect of $z_2$ appears at $\e^4$ order. Up to  $\e^2$ order, the entangling surface area reads
\bea		\la{res:minsurdisk}
A_0  &=& \frac{l^2 \Omega _2}{2 \epsilon ^2} - \frac{\Omega _2 }{2} \log
   \left(\frac{l}{\epsilon }\right) + {\cal O}(1)  , \nn
{A}_{20}  &=& \frac{1}{6 }  \k^2 m_q^2  l^2 \Omega _2
   \log \left(m_q \epsilon  \right)  + {\cal O}(1) , \nn
{A}_{22} &=&  \frac{1}{12}    \kappa^2 m_q^2   l^2 \Omega _2  
\log \left(m_q \epsilon\right)  + {\cal O}(1)   .
\eea
Above $A_0 $ is associated with the entanglement entropy of the undeformed CFT
\be
S_0   \equiv  \fr{2 \pi A_0}{\k^2} 
= \frac{\pi  l^2 \Omega _2}{\kappa ^2 \epsilon ^2} 
- \frac{\pi  \Omega _2 }{ \kappa ^2} \log \left(\frac{ l} {\epsilon } \right) + {\cal O} (1) .
\ee
After setting $\d = m_q \e^2$ and $m_q = \l \m$ as mentioned before, ${A}_{20}$ caused by the metric deformation is in agreement with the result in \cite{Hung:2011ta} for $d=4$ and $\a=1$ (or $m=2$). Above ${A}_{22}$ shows a new logarithmic contribution generated by the second order minimal surface deformation. 
As a consequence, the final logarithmic correction caused by the relevant deformation becomes
\be
\d S = \fr{2 \pi}{\k^2} \ls {A}_{20} + {A}_{22} \rs  =  
\frac{\pi  }{2 }  m_q^2  A_{\S} \log \left(m_q \epsilon  \right) ,
\ee
where the area of the entangling surface is given by $A_{\S} = l^2 \O_2$. 
This result is exact since there is no more logarithmic correction from higher order deformations.

\subsection{ Logarithmic correction to a three-dimensional CFT}

There is no logarithmic term related to the central charge in an odd dimensional theory, whereas the additional logarithmic correction caused by a relevant deformation exists even in an odd dimension. In this section, we will study such an additional logarithmic correction in an odd-dimensional strongly interacting CFT. To do so, let's consider a holographic dual of a three-dimensional CFT. For obtaining a logarithmic correction, we take into account a scalar field with the mass, $m_{\ph}^2 = -5/4$. Then, the resulting metric and the scalar field satisfying all equations of motion near the boundary are given by
\bea
ds^2 &=& \frac{1}{z^2} \lb f(z) \ls - dt^2 + d x_1^2  + d x_2^2\rs  + dz^2 \frac{}{} \rb , \nn
\ph(z) &=& d_1 z^{1/2} -\frac{3}{16} d_1^3 \kappa ^2 z^{3/2} + \left(d_2-\frac{27}{256} d_1^5 \kappa ^4 \log z \right)  z^{5/2} + \cdots ,
\eea
with
\bea
f(z) &=& 1 -\frac{1}{4} d_1^2 \kappa ^2 z +\frac{13}{128} d_1^4 \kappa ^4 z^2 - z^3 
\left(\frac{113 d_1^6 \kappa ^6}{3072}+\frac{5}{18} d_1 d_2  \kappa ^2
-\frac{15}{512} d_1^6 \kappa ^6 \log z \right) + \cdots  ,
\eea
where $d_1$ and $d_2$ are two undetermined integration constants. The dual operator of the scalar field has the conformal dimension $\D=5/2$. Following the AdS/CFT correspondence, $d_1$ and $d_2$ can be interpreted as the source
and the vev of the dual operator respectively.

When we put a subsystem into a strip represented as
\be
- \fr{l}{2} \le x_1 =x \le \fr{l}{2}  \quad {\rm and}  \quad - \fr{L}{2} \le x_2  \le \fr{L}{2} ,
\ee
the turning point is given in terms of $l$ 
\be
z_* =  l  \ls \frac{3 \Gamma \left(\frac{5}{4}\right)}{2 \sqrt{\pi } \Gamma \left(\frac{7}{4}\right)} 
- \frac{27  \kappa ^2 \Gamma
   \left(\frac{5}{4}\right)^3 (-2 K(-1)-1+2 E(-1))}{64 \pi ^{3/2} \Gamma \left(\frac{7}{4}\right)^3}\  d_1^2  l   + {\cal O} \ls d_1^3 l^{3/2} \rs  \rs ,
\ee 
and the area of the minimal surface reads
\bea 		\la{res:area3}
A &=&\frac{2 L}{\epsilon } -\frac{4 \pi  L \Gamma \left(\frac{7}{4}\right)^2}{9 \Gamma
   \left(\frac{5}{4}\right)^2} \ \fr{1}{l} \nn
   && - \frac{d_1^2 l \kappa ^2 L}{8}   \left(2 \log l -2 \log \epsilon +1+\log 2
   +2 \log \left[ \frac{3 \Gamma \left(\frac{5}{4}\right)}{2
   \sqrt{\pi } \Gamma \left(\frac{7}{4}\right)}\right)   \right]  \fr{1}{l}+ {\cal O} \ls d_1^3  l^{3/2} \rs ,
\eea
where $E(x)$ and $K(x)$ imply two complete elliptic integrals, $ E(-1)=1.9101$ and $K(-1)=1.31103$. The first line in \eq{res:area3} is the minimal surface area of the undeformed theory, while the remaining is the contribution from the relevant deformation. This result shows that, as expected, the operator with $\D=5/2$ in a three-dimensional CFT gives rise to a  logarithmic correction 
\be
\d S = \frac{\pi}{4} d_1^2  A_{\S} \log \ls d_1^2 \epsilon  \rs ,
\ee
where $A_{\S} = 2 L$ is the area of the entangling surface.

Now, let us turn into a subsystem in a disk. The minimal surface bordering on the entangling surface has the following induced metric
\be
ds_{in}^2 =  \frac{1}{z^2} \lb \ls f(z) + z'^2 \rs  d \r^2  + f(z)  \r^2 d \th^2  \rb .
\ee
If the radius of the entangling surface is denoted by $l$, the area of the minimal surface is reduced to
\be			\la{act:minimaldisk}
A =  2 \pi \int_{0}^{l -\r_*} d \r \ \fr{\r \sqrt{f(z)} }{z^2} \sqrt{f(z) + z'^2 } ,
\ee
where $\r_*$ denotes a UV cutoff in the $\r$ coordinate which is related to the UV cutoff in the $z$-direction. Its equation of motion reads
\be
0 = 2 f \rho  z z'' +2 z z'^3 + 4 f^2 \rho + \left(4 f \rho -3 \rho  z f'\right)  z'^2  -2 f \rho  z f'+2 f z z'  .
\ee
In order to solve this equation perturbatively in the $ d_1^2 l \ll $ limit, let's take the following ansatz
\be
z(\r) = z_0 (\r) + d_1 l^{1/2}  \ z_1 (\r)+  d_1^2 l  \ z_2 (\r) + {\cal O} (d_1^3 l^{3/2}) .
\ee
The first term is the solution of the undeformed theory which is again given by $z_0  = \sqrt{l^2 - \r^2}$. 
The first correction, $z_1(\r)$, vanishes due to the invariance under $d_1 \to - d_1$ as mentioned before.
At $d_1^2 l$ order, the second correction, $z_2$ satisfies 
\be
0 = z_2''+ \frac{\left(l^2-2 \rho ^2\right) }{\rho  (l^2 -\rho^2 )} z_2' 
-\frac{2 l^2 }{(l^2 -\rho^2 )^2} z_2
+\frac{\kappa ^2 \left(3 \rho ^2-2 l^2\right)}{8 l (l^2 -\rho^2 )} .
\ee
This differential equation yields the following exact solution 
\bea
z_2&=& \fr{d_3}{\sqrt{l^2-\rho ^2}} +  d_4 + \frac{d_4  l \lb 
\log \rho -\log \left( l^2 + l \sqrt{l^2-\rho
   ^2}\right)  \rb}{\sqrt{l^2-\rho ^2}}
+\frac{3 \kappa ^2 l^2}{32 \sqrt{l^2-\rho ^2}}+\frac{\kappa ^2 \rho ^2}{16 l}-\frac{3 \kappa ^2 l}{16} \nn
&& + \frac{\kappa ^2 l^2 \log l \log \rho }{8 \sqrt{l^2-\rho ^2}}
+\frac{1}{16} \kappa ^2 l \log \left(l^2-\rho^2\right) 
+\frac{\kappa ^2 l^2 \log \left(l^2-\rho ^2\right) \log \left(\frac{l}{\sqrt{l^2-\rho ^2}+l}\right)}{16 \sqrt{l^2-\rho ^2}} \nn
   && -\frac{\kappa ^2 l^2 \text{Li}_2\left(\frac{\rho ^2}{l^2}\right)}{32 \sqrt{l^2-\rho ^2}}-\frac{\kappa ^2
   l^2 \text{Li}_2\left(-\frac{\sqrt{l^2-\rho ^2}}{l}\right)}{8 \sqrt{l^2-\rho ^2}} ,
\eea
where $\text{Li}_2 (z)= \sum_{k=1}^{\infty} \fr{z^k}{k^n}$ indicates a polylogarithm function. Two required boundary conditions determine  two integration constants to be
\bea
d_3 &=&  \frac{1}{192} \kappa ^2 l^2 \left(-48 \log ^2 l +\pi ^2-18\right) , \nn
d_4 &=&  -\frac{1}{8} \kappa ^2 l \log l . 
\eea

Substituting these solutions into \eq{act:minimaldisk}, the minimal surface area near the asymptotic boundary is expanded into 
\be   \la{res:disk3}
A =   \fr{2 \pi l}{\e} +  \ls \fr{1}{4}   + \fr{1}{12} \rs  \pi l \k^2 d_1^2  \log  \ls d_1^2 \e \rs   + {\cal O} (1) ,
\ee
where $\r_* = \fr{\e^2}{2 l}$ was used. Above the first comes from the undeformed theory, whereas the logarithmic corrections are caused by the metric and minimal surface deformation. The first logarithmic correction is due to the metric deformation which is in agreement with the result in \cite{Hung:2011ta} when identifying $\d = d_1^2 \e^2$.
The second is the contribution from the second order minimal surface deformation. 
Recalling that the circumference of the disk is given by $A_{\S} = 2 \pi l$, the exact logarithmic correction in the deformed CFT is given by
\be
\d S =  \fr{\pi}{3} d_1^2 A_{\S} \log \ls d_1^2 \e \rs .
\ee


\section{Discussion}

In a strongly interacting CFT which has an AdS dual gravity, we have investigated logarithmic corrections caused by relevant deformation operators. To do so, we considered a massive scalar field dual to an operator with a conformal dimension $\D=\fr{d+2}{2}$ and studied the dual geometry including its gravitational backreaction. In general, a relevant operator alters both the metric and minimal surfaces which usually cause the change of the holographic entanglement entropy. It has been well studied in \cite{Hung:2011ta} how the metric deformation gives rise to a logarithmic correction. In this work, we have focused on a logarithmic term caused by the minimal surface. 
Through the explicit calculation of the entanglement entropy in  a disk-shaped region, we showed for $n=1$ that  the second order minimal surface deformation leads to an additional logarithmic correction. In this case, there is no more logarithmic term caused by higher order metric and minimal surface deformations, our results are exact. 

For a general $n$, in addition to the logarithmic term caused by the metric deformation we can expect an additional logarithmic term generated by the 
$2n$-th order minimal surface deformation. Furthermore, it seems to be possible to get more general logarithmic corrections from the combinations of the metric and minimal surface deformations. We hope to report some results on this issue in future works.

\vspace{1cm}

{\bf Acknowledgement} 

We thank to T. Nishioka for valuable discussion. C. Park was supported by Basic Science Research Program through the National Research Foundation of Korea funded by the Ministry of Education (NRF-2013R1A1A2A10057490) and also by the National Research Foundation of Korea grant funded by the Korea government (MSIP) (2014R1A2A1A01002306). C. Park also acknowledges the Korea Ministry of Education, Science and Technology, Gyeongsangbuk-Do and Pohang City.  

\vspace{1cm}


\end{document}